\documentclass[pre,aps,showpacs,twocolumn,floatfix]{revtex4}
\usepackage[dvips]{graphicx}
\usepackage{amsmath}
\usepackage{amsfonts}
\usepackage{amssymb}

\newcommand{\be}{\begin{equation}}
\newcommand{\ee}{\end{equation}}
\newcommand{\bea}{\begin{eqnarray}}
\newcommand{\eea}{\end{eqnarray}}
\newcommand{\ba}{\begin{array}}
\newcommand{\ea}{\end{array}}

\begin{document}

%
%
\title{Resonant activation driven by strongly non-Gaussian noises.}

\author{Bart{\l}omiej \surname{Dybiec}}
\email{bartek@th.if.uj.edu.pl}

\author{Ewa \surname{Gudowska--Nowak}}
\email{gudowska@th.if.uj.edu.pl}
\affiliation{Marian~Smoluchowski Institute of Physics,\\
 Jagellonian University, Reymonta~4, 30--059~Krak\'ow, Poland}

\date{\today}

\begin{abstract}

The constructive role of non-Gaussian random fluctuations is studied in the
context of the passage over the dichotomously switching potential barrier.
Our attention focuses on the interplay of the effects of independent sources of
fluctuations: an additive stable noise representing non-equilibrium external
random force acting on the system and a fluctuating barrier.
In particular, the influence of the structure of stable noises on the
mean escape time and on the phenomenon of resonant activation (RA) is
investigated.
By use of the numerical Monte Carlo method it is documented that the suitable choice
of the barrier switching rate and random external fields may produce
 resonant phenomenon leading to the enhancement of the kinetics and the
 shortest, most efficient reaction time.

\end{abstract}


\maketitle


\section{Introduction}

The fluctuating barrier problem seems to be of relevance especially for the
kinetics of biophysical systems \cite{doe,ful,astumian,ewa} whose time-evolution is governed by coupling to
protein molecules undergoing constant conformational transitions.
In particular, it is by now well established that membrane proteins that conduct many chemical
 and
physical processes can exist in several conformational states whose distinct electrical
properties may be responsible for the efficiency of ionic transport through biological
channels. In such cases, the assumption of random fluctuations in the membrane potential has
been shown \cite{ful,bez} to be sufficient to explain experimental data on active ionic flow and resonance
between internal fluctuations and external driving AC field. The enhancement or optimisation of
the action of the
external regular field by stochastic fluctuations gained the term {\it stochastic resonance} (SR) and
became an issue of vivid experimental and theoretical studies \cite{gammaitoni} over the past 20
years.
Although the SR phenomenon is commonly described as a cooperative effect in which small periodic
influence entrains external random noise, similar constructive effects of noises under nonequilibrium
constraints can be expected in ``ratchet'' systems \cite{reimann} where spatially uniform mean-zero symmetrical (random
or periodic) time
dependent forces interact with an underlying spatial anisotropy to induce motion. Another
resonant-type, noise-induced behaviour \cite{doe,boguna,marchesoni,iwaniszewski} is resonant activation (RA), the phenomenon in which an
optimal fluctuation rate exists which minimizes the mean first passage time for the
escape over the fluctuating barrier.
The generic theoretical models analysing the aforementioned resonant phenomena are usually based on a
Langevin equation approach assuming the overdamped limit \cite{gammaitoni,doe,dybiec1,dybiec2}. Accordingly, the influence
of the external thermal bath of the surroundings on a Brownian particle is described in such an
equation by time-dependent random force which is commonly assumed to be represented by a white Gaussian
noise. That postulate is compatible with the assumption
of a short correlation time of fluctuations, much shorter than the time-scale of
the macroscopic
motion and assumes that weak interactions with the bath lead to independent
random variations of the parameter describing the motion.
In more formal, mathematical terms Gaussianity of the state-variable fluctuations
is a consequence of the Central Limit Theorem which states that normalized
sum of independent and identically distributed ({\it i.i.d}) random variables with finite
variance converges to the Gaussian probability distribution. If, however,
after random collisions jump lengths are ruled by broad distributions leading to
the divergence of the second moment, the statistics of the process changes
significantly. The existence of the limiting distribution is then guaranteed
by the generalized L\'evy-Gnedenko \cite{gnedenko} limit theorem. According to the latter,
normalized sums of independent, identically distributed random variables with infinite variance converge
in distribution to the L\'evy statistics. At the level of the Langevin equation,
L\'evy noises are generalization of the Brownian motion and describe results of
strong collisions between the test particle and the surrounding environment. In
this sense, they lead to different models of the bath that go beyond a standard
``close-to-equilibrium'' Gaussian description~\cite{dietlevsen,cohen,garbaczewski,bartek}.

In this communication we present numerical results for the mean first
passage time over a
fluctuating barrier for the model system with a linear potential
subject to Markovian dichotomous fluctuations and additive L\'evy noises.
 We report here on a new phenomenon of reappearance of RA in a model system whose pattern
 of the most efficient kinetics has been smeared out and destroyed by the presence of driving
 $\alpha$-stable noise. In particular, we demonstrate that a congruent tuning of noise asymmetry
 with its stability index responsible for scaling properties of the noise increments can optimize the passage over the conformationally changing
 barrier.

\section{The Model}
An overdamped Brownian particle is moving in a potential field
between absorbing ($x=1$) and reflecting ($x=0$) boundaries,
 in the presence of noise that modulates the barrier height.
Time evolution of a state variable $x(t)$ is described in terms of the
generalized Langevin equation
\begin{equation}
\frac{dx}{dt}  =  -V'(x)+g\eta(t)+\sqrt{2}\zeta(t) =-V_\pm^{'}(x)+\sqrt{2}\zeta(t),
\label{lang}
\end{equation}
where prime means differentiation over $x$, $\zeta(t)$ is a white L\'evy process originating
from the contact with non-equilibrated bath and
$\eta(t)$ stands for a Markovian dichotomous noise of intensity $g$ taking one of
two possible values $\pm 1$.
Autocorrelation of the dichotomous noise is set to
$\langle (\eta(t)-\langle\eta\rangle)(\eta(t')-
\langle\eta\rangle)\rangle=
\exp(-2\gamma |t-t'|)$.
For simplicity, throughout the paper a particle mass, a friction coefficient and the
 Boltzman constant are all set to 1. The time-dependent potential
$V_\pm(x)$ is assumed to be linear with the barrier switching between two configurations
with an average rate $\gamma$
\begin{equation}
V_\pm(x)=H_\pm x, \qquad g=\frac{H_{-}-H_{+}}{2}.
\end{equation}
Both $\zeta$ and $\eta$ noises are assumed to be statistically independent.

The initial condition for Eq.~(\ref{lang}) is $x(0)=0$, so that
 initially particle is located at the reflecting boundary with
equal choices of finding a potential barrier in any of two possible configurations $(\pm)$.
The quantity of interest is the mean first passage time (MFPT),
the average time which particle spends in the system before it becomes absorbed.
In the approach applied herein information on the MFPT is drawn
from the statistics of numerically generated trajectories satisfying
 the generalized Langevin equation (\ref{lang}) and
the examined MFPT is estimated as a first moment
of the distribution of first passage times (FPT) obtained
from the ensemble of simulated realizations of the stochastic process in question.
More precisely, frequency of FPT is calculated as frequency of the time-duration $\tau$ of the events (trajectories)
that have initially started
 from $x=0$ and reached $x=1$ for the first time. The MFPT is estimated as $\langle \tau \rangle$ over
 the frequency distribution.
Despite the fact that in Eq.~(\ref{lang}) $x$ itself is a random variable distributed according to some unknown
stable distribution, the FPTs are distributed according to some unknown probability distribution
possessing all moments.

\section{Results of Simulations}

Generally, for systems driven by Gaussian noises, noise-induced phenomena are studied
solely in terms of
$\sigma^2$, the intensity of the white noise \cite{gammaitoni,lefever}. Thermodynamic
definition of this noise-source relates $\sigma^2$ to the system temperature. It is not so, however,
for dynamic systems perturbed by L\'evy stable noises, for which an appropriate choice of a control
parameter is less obvious.
It is caused by the fact that stable distributions are characterized by a four-parameter
 family:
the stability index $\alpha$ ($\alpha\in(0,2]$), skewness parameter $\beta$ ($\beta\in[-1,1]$),
 the
shift $\mu$ ($\mu\in\mathbb{R}$) and scaling $\sigma$ ($\sigma\in\mathbb{R^+}$)
parameters,
 respectively (for the definition of L\'evy measure, see Appendix A).
The PDFs of stable distributions demonstrate asymptotic power-law behaviour
according to
$L(\zeta)\approx |\zeta|^{-1-\alpha}$~\cite{gnedenko,weron}. The parameter $\mu$ controls
the position of the PDF modal value, whereas $\sigma$, similarly to the Gaussian case,
scales the width of the distribution.
At the first glance, the parameter $\sigma$ is somehow connected to the system temperature,
although the relation is more subtle since stable additive noises can be expected in
nonequilibrium situations where the definition of temperature looses its common sense
\cite{bartek,dietlevsen}.

For the purpose of simulations the value of $\mu$ has been arbitrary set to 0 and $\sigma=1/\sqrt{2}$.
Such a choice of $\mu$ and $\sigma$ reconstructs standard normal distribution $N(0,1)$ for
 $\alpha=2$, therefore allowing for comparison of the
 results obtained in this study with the former ones ~\cite{doe,dybiec1,dybiec2}, where the Gaussian measures of the
 $\zeta$ noise have been used.
Remaining parameters ($\alpha$ and $\beta$) have taken values from the allowed range with a
lower limit set up for the $\alpha$ value ($\alpha<.2$ have not been investigated).
The RA phenomenon has been examined for various barrier setups: The potential barrier
 has been switching between different heights $H_+$ and $H_-$ and as
 values of $H_\pm$ the sets $\pm8$; (0, 8); and (4, 8) have been chosen.
They correspond to the changing of the potential barrier between the barrier
 and the well ($H_\pm=\pm8$),
the barrier and ``no barrier'' ($H_+=8$, $H_-=0$) situations and to two barriers with
different heights ($H_+=8$, $H_-=4$).

\begin{figure}[!ht]
\begin{center}
\includegraphics[angle=0, width=7.0cm, height=9.0cm]{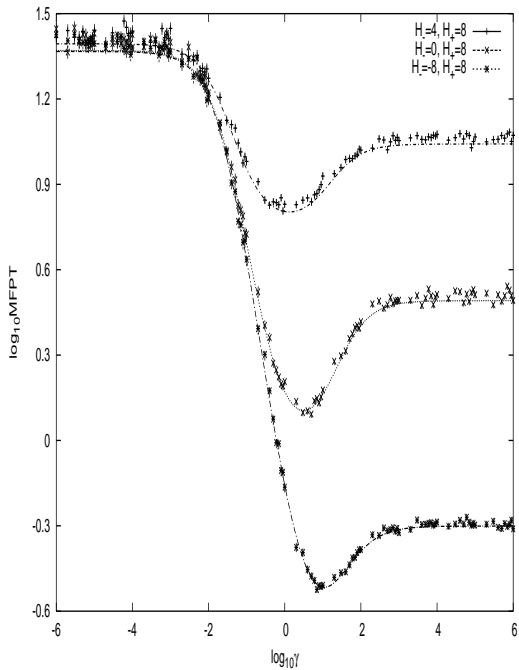}
\caption{MFPT$(\gamma)$ for linear potential barriers switching between different heights $H_\pm$:
$(+)\;H_+=8,\;H_-=4$; $(\times)\;H_+=8,\;H_-=0$; $(\ast)\;H_+=8,\;H_-=-8$.
The driving noise is L\'evy stable noise with $\alpha=2$, i.e. Gaussian noise.
Solid lines represent exact result constructed by direct integration of the backward Fokker--Planck equation.
Numerical results were obtained by use of Monte Carlo simulation of Eq.~(\ref{lang}) with time step $dt=10^{-4}$
and averaged over $N=10^3$ realizations. Error bars represent deviation from the mean and usually remain within the symbol size.}
\label{gausplot}
\end{center}
\end{figure}

In order to test the implemented numerical procedures (cf. Appendix A), MC solutions of Eq.~(\ref{lang}) for $\alpha=2$
have been compared with the exact solutions to the backward Fokker Planck Equation~\cite{doe,dybiec1}.
The numerical results (see Fig.~\ref{gausplot}) have shown a perfect agreement with the exact solutions.

\begin{figure*}[!ht]
\begin{center}
\includegraphics[angle=0, width=14.0cm, height=9.5cm]{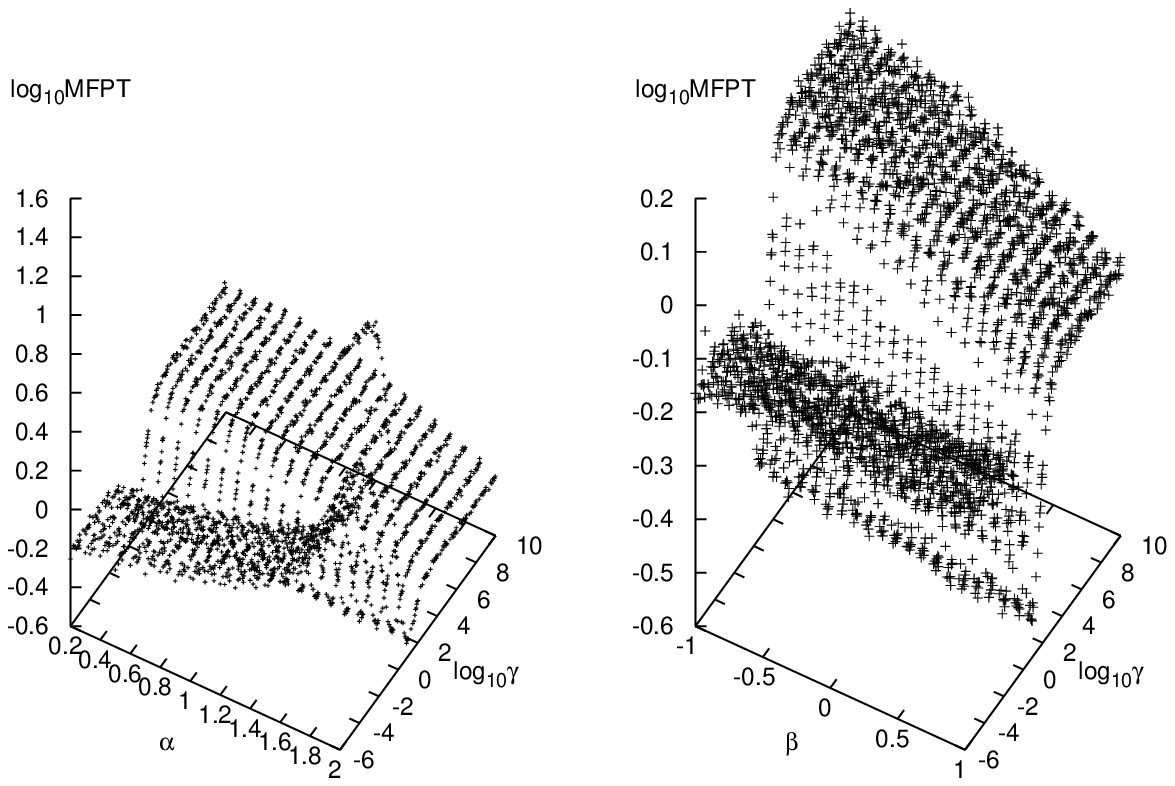}
\caption{MFPT$(\alpha,\gamma)$ for $\beta=0$ (left panel) and MFPT$(\beta,\gamma)$ for $\alpha=0.5$
(right panel) for the linear potential barrier switching between $H_\pm=\pm8$.
The results were calculated by direct integration of Eq.~(\ref{lang}) with the time step $dt=10^{-4}$
and averaged over $N=10^3$ realizations.}
\label{alphabetam88plot}
\end{center}
\end{figure*}

\begin{figure*}[!ht]
\begin{center}
\includegraphics[angle=0, width=14.0cm, height=9.5cm]{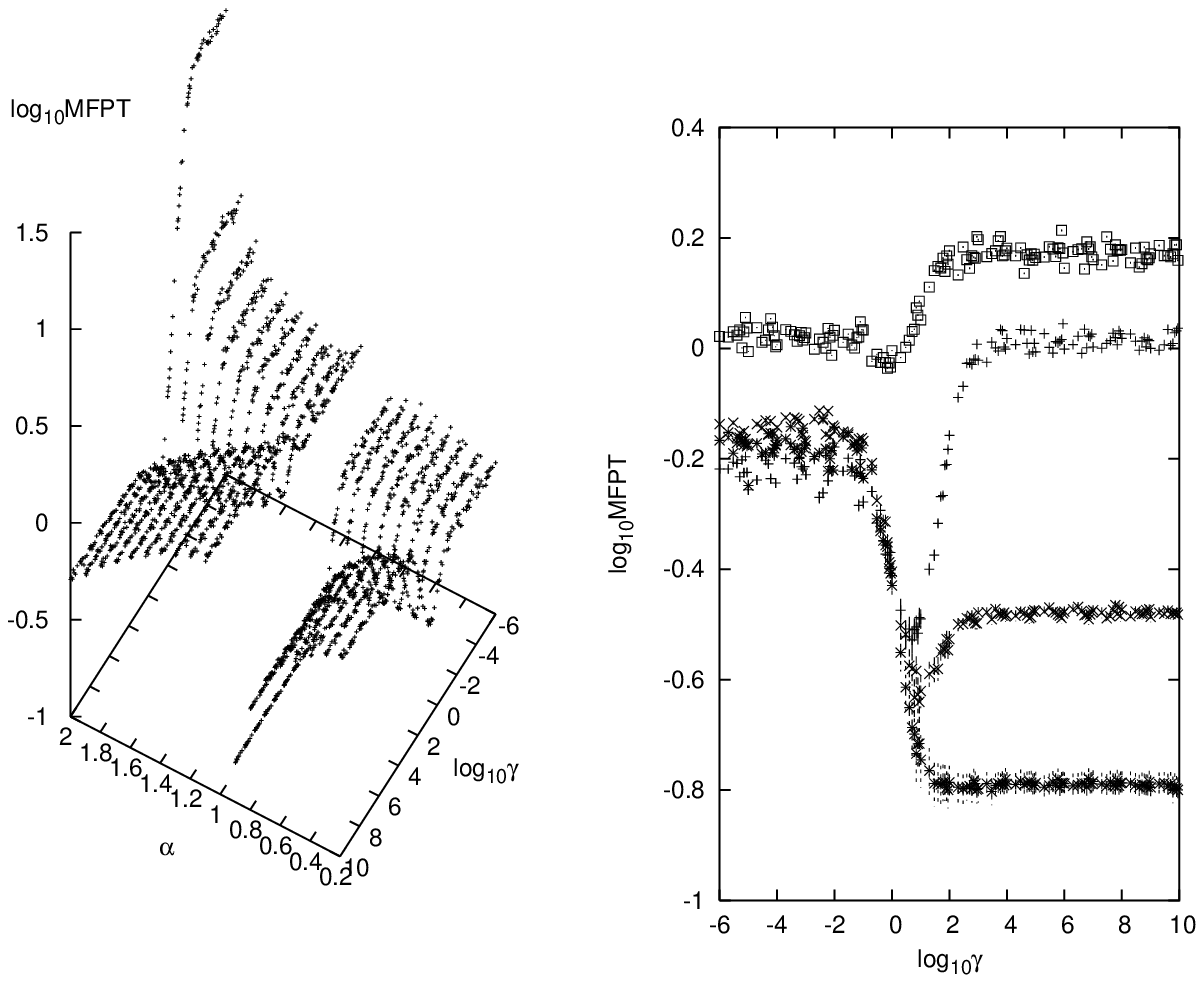}
\caption{MFPT$(\alpha,\gamma)$ for $H_\pm=\pm8$ and $\beta=1$ (left panel) and
sample cross-sections MFPT$(\gamma)$ for various $\alpha$:
$(+)\;\alpha=0.2$; $(\times)\;\alpha=0.8$; $(\ast)\;\alpha=0.9$;  $(\square)\;\alpha=1.1$ (right panel).
The results were calculated by direct integration of Eq.~(\ref{lang}) with the time step $dt=10^{-4}$
and averaged over $N=10^3$ realizations. Error bars represent deviation from the mean.}
\label{alphaskewedm88plot}
\end{center}
\end{figure*}

\begin{figure*}[!ht]
\begin{center}
\includegraphics[angle=0, width=14.0cm, height=9.5cm]{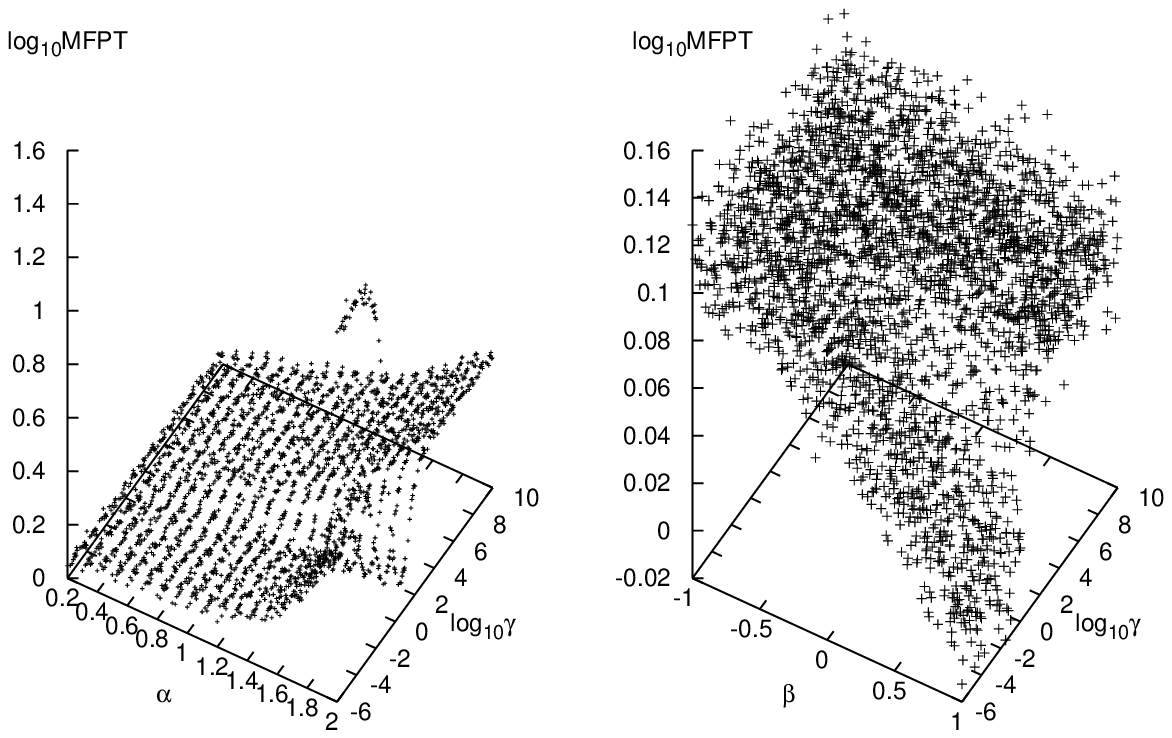}
\caption{MFPT$(\alpha,\gamma)$ for $\beta=0$ (left panel) and MFPT$(\beta,\gamma)$ for $\alpha=0.5$
(right panel) for the linear potential barrier switching between $H_+=8,\;H_-=0$.
The results were calculated by direct integration of Eq.~(\ref{lang}) with the time step $dt=10^{-4}$
and averaged over $N=10^3$ realizations.}
\label{alphabeta08plot}
\end{center}
\end{figure*}

\begin{figure*}[!ht]
\begin{center}
\includegraphics[angle=0, width=14.0cm, height=9.5cm]{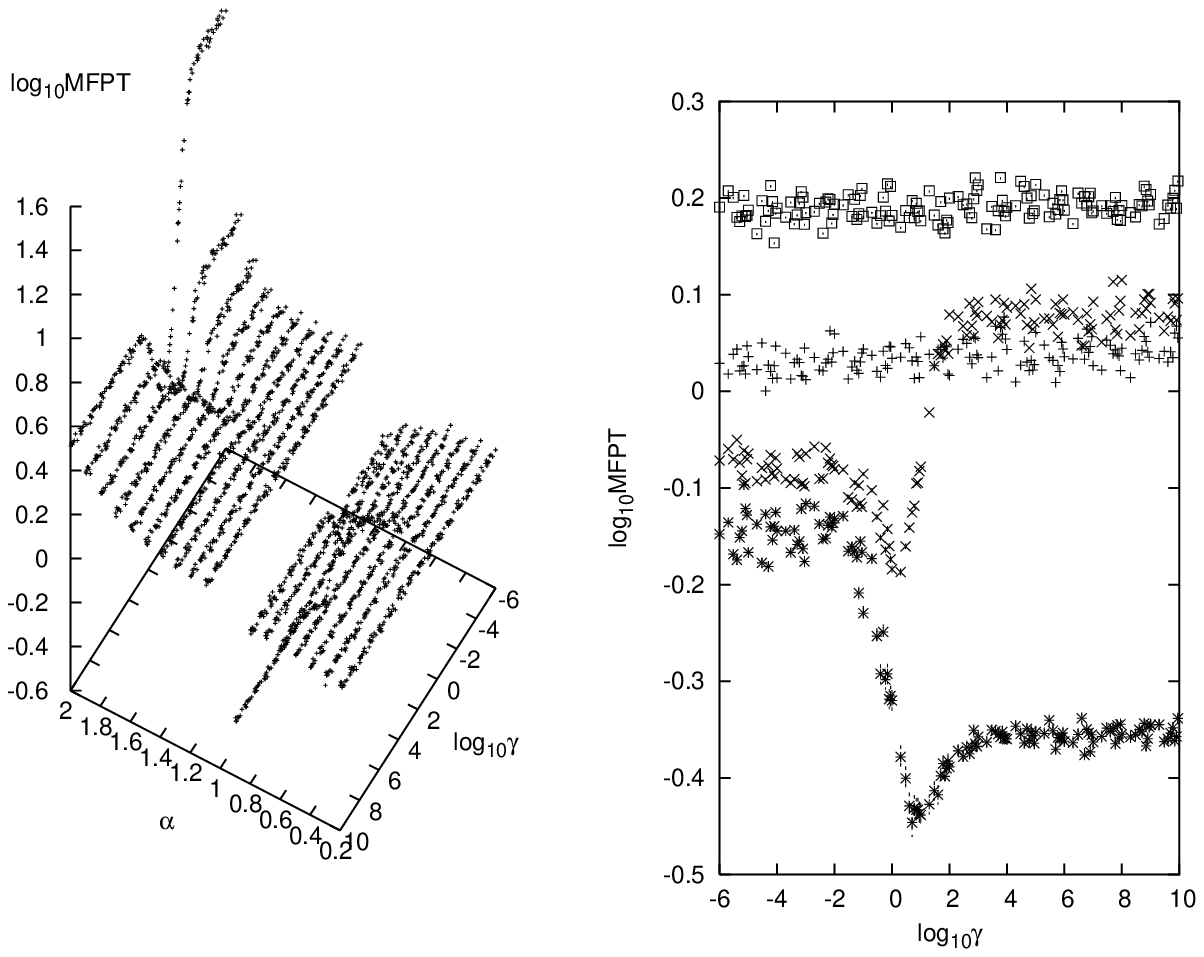}
\caption{MFPT$(\alpha,\gamma)$ for $H_+=8,\;H_-=0$ and $\beta=1$ (left panel) and
sample cross-sections MFPT$(\gamma)$ for various $\alpha$:
$(+)\;\alpha=0.2$; $(\times)\;\alpha=0.8$; $(\ast)\;\alpha=0.9$;  $(\square)\;\alpha=1.1$ (right panel).
The results were calculated by direct integration of Eq.~(\ref{lang}) with the time step $dt=10^{-4}$
and averaged over $N=10^3$ realizations. Error bars represent deviation from the mean.}
\label{alphaskewed08plot}
\end{center}
\end{figure*}

\begin{figure*}[!ht]
\begin{center}
\includegraphics[angle=0, width=14.0cm, height=9.5cm]{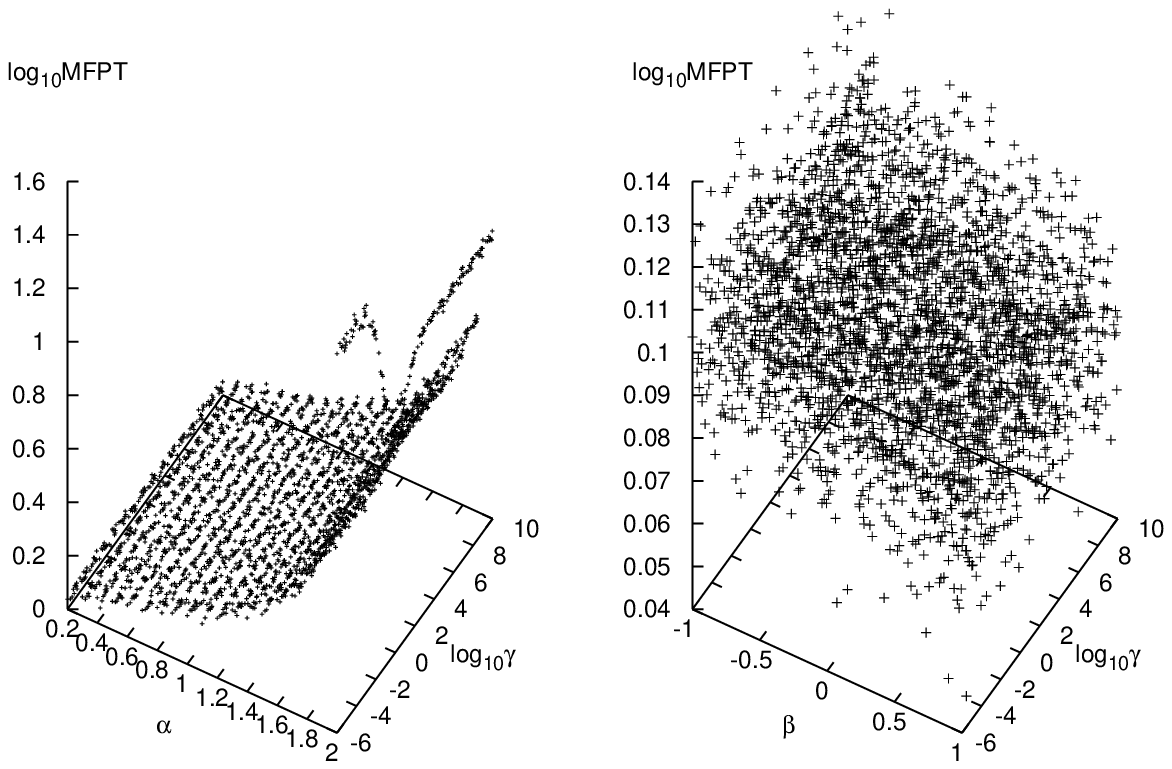}
\caption{MFPT$(\alpha,\gamma)$ for $\beta=0$ (left panel) and MFPT$(\beta,\gamma)$ for $\alpha=0.5$
(right panel) for the linear potential barrier switching between $H_+=8,\;H_-=4$.
The results were calculated by direct integration of Eq.~(\ref{lang}) with the time step $dt=10^{-4}$
and averaged over $N=10^3$ realizations.}
\label{alphabeta48plot}
\end{center}
\end{figure*}

\begin{figure*}[!ht]
\begin{center}
\includegraphics[angle=0, width=14.0cm, height=9.5cm]{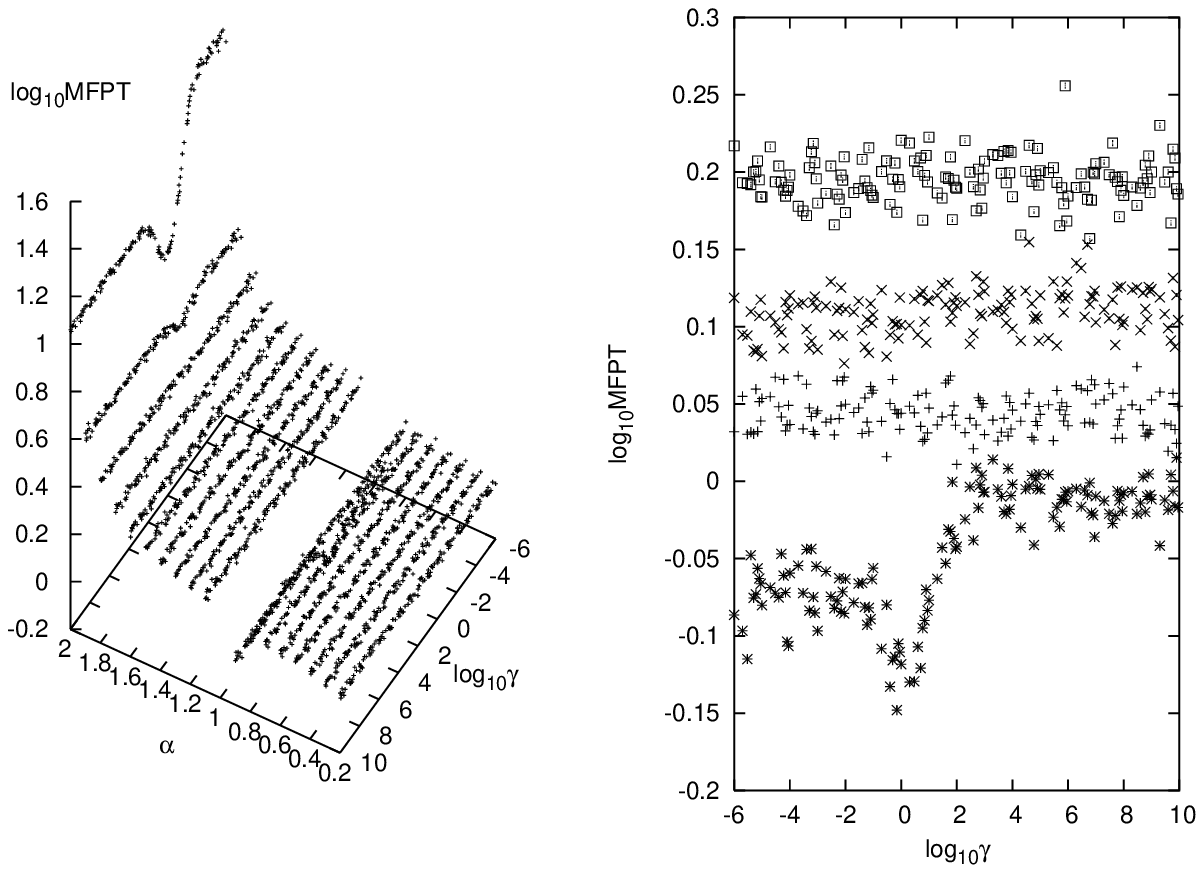}
\caption{MFPT$(\alpha,\gamma)$ for $H_+=8,\;H_-=4$ and $\beta=1$ (left panel)
and sample cross-sections MFPT$(\gamma)$ for various $\alpha$:
$(+)\;\alpha=0.2$; $(\times)\;\alpha=0.8$; $(\ast)\;\alpha=0.9$;  $(\square)\;\alpha=1.1$ (right panel).
The results were calculated by direct integration of Eq.~(\ref{lang}) with the time step $dt=10^{-4}$
and averaged over $N=10^3$ realizations. Error bars represent deviation from the mean.}
\label{alphaskewed48plot}
\end{center}
\end{figure*}

Subsequent figures demonstrate results of the analysis for driving stable noises with various values of the
$(\alpha, \beta)$ parameters. For the sake of clarity of the comparison, they have been grouped in three
classes corresponding to different scenarios of barrier-alteration.
The first group (Figs.~\ref{alphabetam88plot}
and~\ref{alphaskewedm88plot}) relates to the potential barrier switching between
 $H_\pm=\pm8$, Figs.~\ref{alphabeta08plot} and~\ref{alphaskewed08plot}
 describe numerical results obtained for the barrier switching between
 $H_+=8$ and $H_-=0$ and, eventually, Figs.~\ref{alphabeta48plot}
 and~\ref{alphaskewed48plot}
 refer to simulations performed
for the potential barrier switching between two different barriers,
 i.e. $H_+=8,H_-=4$.
In each class, in the first (left) plot (cf. Figs.~\ref{alphabetam88plot},
~\ref{alphabeta08plot} and~\ref{alphabeta48plot}) the results for fixed
 $\beta$ ($\beta=0$), i.e. for the symmetric stable noises are displayed
 while the right panel presents results for a fixed value of
 $\alpha$ ($\alpha=0.5$), thus
 demonstrating the effect of the asymmetry of the driving noise.
The second set of plots in each class (cf. Figs.~\ref{alphaskewedm88plot},~\ref{alphaskewed08plot}
and~\ref{alphaskewed48plot}) presents results for
the
MFPT($\alpha,\gamma$) for a fixed $\beta$ parameter ($\beta=1$) while a corresponding
right panel displays sample cross-sections of the MFPT($\alpha,\gamma$) surface.

Closer examination of Figs.~\ref{alphabetam88plot} and~\ref{alphaskewedm88plot}
allows to conclude that a typical, non-monotonic character of the MFPT curve is
preserved for all values of the $\alpha$ parameter.
The only visible changes are registered in the asymptotic behavior
and in the depth of the MFPTs curves.
In particular, the values of MFPTs for the Gaussian case ($\alpha=2$) are
higher than for other values of $\alpha$ thus demonstrating that the heavier
tails in PDFs (and therefore higher probability of experiencing larger
fluctuations) of driving random forces facilitate the transport over the
fluctuating barrier.
Furthermore, as it can be seen in the right panel of
 Fig.~\ref{alphabetam88plot}, for $\alpha=0.5$ the RA phenomenon is observed for
 all values $\beta$.

In Fig.~\ref{alphaskewedm88plot} MFPT$(\alpha,\gamma)$
surfaces for asymmetric stable noises ($\beta=1$) are presented.
Interestingly, the MFPT$(\alpha,\gamma)$ surface for asymmetric ($\beta=1$)
driving noise displays an unexpected behavior.
Namely, a gradual change of the $\alpha$ parameter (from $\alpha=0.2$ to $\alpha=1.1$)
results in suppression of the RA phenomenon that disappears for values of $\alpha\approx
1$ but becomes further recovered for $\alpha>1$ approaching $\alpha$
representative for the Gaussian case. By comparison (cf. Fig.~\ref{alphabeta08plot}, left panel),
 in the cases when the dynamics flickers between the motion over the erected
 potential barrier and a free (symmetric) L\'evy flight, the characteristic resonant shape of
the MFPT curve
disappears with a decrease of the stability index $\alpha$, i.e. noises with heavier
asymptotic tails destroy the RA. Contrary to the $H_\pm=\pm8$ case, for $\alpha=0.5$, the
RA phenomenon is not observed, independently of the value of $\beta$ (cf. Fig.
~\ref{alphabeta08plot}, right panel). However,
for asymmetric stable noises (Fig.~\ref{alphaskewed08plot}) the same kind of behavior
which has been observed for $H_\pm=\pm8$ can be noticed. Here again, it is
possible to detect an optimal $\alpha$-value at which activation processes
are more efficient (they lead to lower values of the MFPT than for all other choices of the
$\alpha$ parameter).
Unlike in the $H_\pm=\pm8$ case, the RA does not take now place for very small values
of $\alpha$ but it reappears for $\alpha\approx1$ when the resonant shape of the MFPT
curve becomes clearly visible.
 Afterwards, for moderate $\alpha$
the RA phenomenon disappears and, as expected, becomes again induced
 for $\alpha=2$.

%
%

%
%

Figs.~\ref{alphabeta48plot} and~\ref{alphaskewed48plot} display results for
$H_+=8,H_-=4$.
As can be observed in the left panel of Fig.~\ref{alphabeta48plot},
 the Gaussian behavior of the MFPT curve
disappears with decrease of the stability index $\alpha$ for symmetric stable noises.
The shape of the MFPT curve typical for the Gaussian case vanishes much faster
than for the $H_+=8,H_-=0$ scenario.
For a heavy tailed ($\alpha=0.5$) external noise driving (right panel of
 Fig.~\ref{alphabeta48plot}) the RA is not visible.

Analogously to the both previously studied barrier setups
with asymmetric stable driving noises (Fig.~\ref{alphaskewed48plot})
there exists an optimal $\alpha$ for which activation processes are
 most efficient. Also, the RA phenomenon is inhibited at low values of $\alpha$,
 reappears for $\alpha\approx1$ and smears
 out for moderate $\alpha$-s before its recovery becomes effective at $\alpha=2$.

%
%

%
%

By inspection of Figs.~\ref{alphaskewedm88plot}, \ref{alphaskewed08plot} and~\ref{alphaskewed48plot}
it can be concluded that the same kind of resonant behavior (described by a
nonmonotonic shape of the MFPT curves) can be observed
not only for equilibrium fluctuations but also for heavy-tailed
nonequilibrium fluctuations
modelled by an asymmetric noise with the stability index $\alpha\approx 1$.
Closer analysis of the MFPT$(\alpha,\gamma)$-surface cross-sections
 shows that the phenomenon disappears for very small $\alpha$ (except of
 the $H_\pm=\pm8$ case), becomes induced for $\alpha\approx 1$, disappears for $\alpha> 1$
and renters again in the Gaussian limit.
Note that the Gaussian cases ($\alpha=2$) have not been plotted in the figures
with cross sections due to the fact that values of the MFPT
for the Gaussian driving noise are significantly higher than for all other stable noises with
smaller values of $\alpha$.
Therefore Fig.~\ref{gausplot} presents separately results of numerical analysis
for $\alpha=2$.

Examination of left panels in Figs.~\ref{alphabetam88plot}, \ref{alphabeta08plot} and~\ref{alphabeta48plot}
allows to see how the typical Gaussian behavior of the MFPT curves changes with the change
of the stability index $\alpha$.
For $H_\pm=\pm8$ only changes observed are visible in the asymptotic behavior of the MFPTs.
For other cases under the study ($H_+=8$, $H_-=0$; $H_+=8$, $H_-=4$),
the typical shape of the MFPT curve disappears with decreasing $\alpha$ and
the RA phenomenon vanishes.
The process is the most rapid for $H_+=8$, $H_-=4$ (cf. left panel of the Fig.~\ref{alphabeta48plot})
while for $H_+=8$, $H_-=0$ it slows down and becomes smoother (cf. left panel of the Fig.~\ref{alphabeta08plot}).

\section{Summary}
We have considered a thermally activated
process that occurs in a system coupled to a non-Gaussian noise
source introduced by a non-equilibrated thermal bath. Another external
stochastic process is assumed to be responsible for dichotomous fluctuations of the potential
barrier which has
been modeled by the linear function with a varying slope.

The applied procedures enable to study the RA phenomena for all possible values
of $\alpha,\beta,\sigma,\mu$ mimicking statistical properties of the driving stable noises.
In particular,
for $\alpha=2$ (and any value of $\beta$), the $\alpha$-stable noise generated
by the recipe given by Eq.~(\ref{recipe1}) is a Gaussian noise and numerically
 constructed results are in a perfect agreement with
the exact results obtained by numerical integration of an appropriate backward
 Fokker-Planck equation~\cite{dybiec1}.
In this case the RA takes place for all barrier setups under consideration.
For $\alpha=0.5,\beta=1$ and $\alpha=1,\beta=0$ formerly studied~\cite{bartek}
 L\'evy-Smirnoff and Cauchy cases were recovered.
Results of numerical analysis presented in this paper agree
with previously constructed solutions~\cite{bartek}.

In the case of driving stable noises, the RA phenomenon is most resistant
 for the $H_\pm=\pm8$ scenario where it is visible for all noises under the study.
For symmetric stable noise it is possible to observe how the Gaussian behavior
of the MFPT curves changes
and the RA phenomenon in the system switching between two barriers
($H_+=8$, $H_-=4$) or barrier and no barrier ($H_+=8$, $H_-=0$) vanishes.
A novel behavior has been registered for nonsymmetric stable noises:
For very small values of the stability index $\alpha$, the RA is not effective
(except of the $H_\pm=\pm8$ case). The phenomenon reappears
by increasing $\alpha$ to $\approx 1$, vanishes for $\alpha>1$
and becomes again observable for $\alpha=2$ (the Gaussian case).
Moreover, in the $\alpha$ space another kind of phenomenon, allowing to choose an
optimal value of $\alpha$
 for which the MFPTs are smaller than for other values of stability index $\alpha$, occurs.
 The fact that the shape of MFPT curve for very heavy-tailed, skewed distributions
 representing non-Gaussian noises arising from the contact with not
equilibrated bath is similar to the MFPT curve in the presence of the Gaussian
fluctuations arising from the contact
with equilibrated thermal bath is very interesting and makes the problem of
recognition of the underlying noise
more complicated.
This indicates that more detailed study of ``resonant phenomena'' in non-equilibrium
situations might be required, as has been recently also argued in other studies
\cite{tessone} relating to the detection of the stochastic resonance in a generic
double well potential under the influence of periodic signal and a colored,
non-Gaussian noise.


\section{$\alpha$-stable Random Variables and $\alpha$-stable L\'evy Motion Processes}

The $\alpha$-stable variables are random variables for which
the sum of random variables is distributed according to the same
distribution as each variable, i.e.
\begin{equation}
aX_1+bX_2\stackrel{\mathrm{d}}{=}cX+d,
\label{def}
\end{equation}
where $\stackrel{\mathrm{d}}{=}$ denotes equality in a distribution sense.
Real constants $c,\;d$ in Eq.~(\ref{def}) allow for rescaling and shifting
of the initial probability distribution.
The characteristic function of the stable distribution can be parameterized
in various ways. In the usually chosen $L_{\alpha,\beta}(\zeta;\sigma,\mu)$
 \cite{janicki,weron} parameterization,
a characteristic function of the L\'evy type variables is given by
\begin{eqnarray}
 \phi(k) & = & \exp\left[ -\sigma^\alpha|k|^\alpha\left( 1-i\beta\mbox{sign}(k)\tan
\frac{\pi\alpha}{2} \right) +i\mu k\right. \nonumber \\
& & - \left. i\beta k\sigma^\alpha\tan \frac{\pi\alpha}{2}\right],\;\;\; \mbox{for}\;\;\alpha\neq 1, \nonumber \\
 \phi(k) & = & \exp\left[ -\sigma|k|\left( 1+i\beta\frac{2}{\pi}\mbox{sign} (k) \ln|k| \right) \right. \nonumber \\
& & +  i\mu k \bigg],\;\;\;\;\;\;\;\;\;\;\;\;\;\;\;\;\; \mbox{for}\;\;\alpha=1,
\label{charakt}
\end{eqnarray}
with
$
\alpha\in(0,2],\;
\beta\in[-1,1],\;
\sigma\in(0,\infty),\;
\mu\in(-\infty,\infty)$ and $\phi(k)$ defined in the Fourier space
\begin{equation}
\phi(k) = \int d\zeta e^{-ik\zeta} L_{\alpha,\beta}(\zeta;\sigma,\mu).
\end{equation}
The above parameterization (\ref{charakt}) is continuous, in the sense that
\begin{equation}
\lim\limits_{\alpha\to1}\sigma^\alpha\left(|k|^\alpha\mbox{sign}
(k) -k \right)\tan \frac{\pi\alpha}{2}
=-\frac{2}{\pi}\sigma|k|\mbox{sign} (k) \ln |k|
\end{equation}
for every $\sigma$ and $k$.
Analytical expressions for stable probability distributions  \linebreak $L_{\alpha,\beta}(\zeta;\sigma,\mu)$
are known only in few cases:
for $\alpha=0.5,\;\beta=1$ resulting distribution is L\'evy-Smirnoff
\begin{eqnarray}
L_{1/2,1}(x;\sigma,\mu) & =  & \left( \frac{\sigma}{2\pi}
\right)^{\frac{1}{2}}(x-\mu)^{-\frac{3}{2}} \nonumber \\
 & \times & \exp\left(-\frac{\sigma}{2(x-\mu)} \right),
\label{smirnoff}
\end{eqnarray}
for $\alpha=1,\;\beta=0$ it results in the Cauchy distribution
\begin{equation}
L_{1,0}(x;\sigma,\mu)=\frac{\sigma}{\pi}\frac{1}{(x-\mu)^2+\sigma^2},
\label{cauchy}
\end{equation}
whereas for $\alpha=2$ with arbitrary $\beta$ the PDF is Gaussian.
Characteristic feature of distributions $L_{\alpha,\beta}(\zeta;\sigma,\mu)$ is existence
of moments of order $\alpha$,
i.e. the integral
 $\int_{-\infty}^{\infty}L_{\alpha,\beta}(\zeta;\sigma,\mu)\zeta^\alpha d\zeta$ is finite.
This statement results in the conclusion that the only stable distribution possessing the
second moment is a Gaussian one; for all other
 values of $\alpha$ the variance of a stable distribution diverges,
and for $\alpha\le1$ also the average value does not exist.

For the purpose of analysis,
the corresponding Langevin equation~(\ref{lang})
has been simulated by use of the appropriate numerical methods.
Position of the particle has been obtained by direct integration of Eq.~(\ref{lang})
\begin{eqnarray}
x(t) & =& -\int_{t_0}^{t}\left[ V'(x(s))-g\eta(s)\right] ds   +  \int_{t_0}^{t} dL_{\alpha,\beta}(s) \nonumber \\
&= & -\int_{t_0}^{t}V_\pm^{'}(x(s))ds+\int_{t_0}^{t}dL_{\alpha,\beta}(s).
\label{lcalka}
\end{eqnarray}
In general~\cite{weron,dietlevsen,janicki}, the $L_{\alpha,\beta}$
 measure in Eq.~(\ref{lcalka}) can be approximated by
\begin{eqnarray}
\int_{t_0}^{t}f(s)dL_{\alpha,\beta}(s) & \approx &
\sum\limits_{i=0}^{N-1}f(i\Delta s)M_{\alpha,\beta}\left([i\Delta s,(i+1)\Delta s)\right) \nonumber \\
& \stackrel{\mathrm{d}}{=} & \sum\limits_{i=0}^{N-1}f(i\Delta s)\Delta s^{1/\alpha}\varsigma_i,
\end{eqnarray}
where $\varsigma_i$ is distributed with the PDF
$L_{\alpha,\beta}(\varsigma;\sigma=\frac{1}{\sqrt{2}},\mu=0)$, $N\Delta s=t-t_0$ and \linebreak $M_{\alpha,\beta}([i\Delta s,(i+1)\Delta s))$ is the measure of the interval $[i\Delta s,(i+1)\Delta s)$.

Random variables $\varsigma$ corresponding to the characteristic function
 (\ref{charakt}) can be generated using the Janicki--Weron algorithm
 \cite{weron}.
For $\alpha\neq1$ their representation is
\begin{eqnarray}
\varsigma  & = & D_{\alpha,\beta,\sigma} \frac{\sin(\alpha(V+C_{\alpha,\beta})) }{
(\cos(V))^{\frac{1}{\alpha}}} \nonumber \\
& \times &
\left[ \frac{\cos(V-\alpha(V+C_{\alpha,\beta}))}{W}
\right]^{\frac{1-\alpha}{\alpha}}
 +  B_{\alpha,\beta,\sigma,\mu},
\label{recipe1}
\end{eqnarray}
with constants $B,C,D$ given by
\begin{equation}
B_{\alpha,\beta,\sigma,\mu}=\mu-\beta\sigma^\alpha\tan(\frac{\pi\alpha}{2}),
\end{equation}
\begin{equation}
C_{\alpha,\beta}=\frac{\arctan\left(\beta\tan(\frac{\pi\alpha}{2})\right)}{
\alpha},
\end{equation}
\begin{equation}
D_{\alpha,\beta,\sigma}=\sigma\left[ \cos\left(
\arctan\left(\beta\tan(\frac{\pi\alpha}{2})\right) \right) \right]^{-\frac{1}{\alpha}}.
\end{equation}
For $\alpha=1$, $\varsigma$ can be obtained from the formula
\begin{eqnarray}
\varsigma  & = & \frac{2\sigma}{\pi} \left[ (\frac{\pi}{2}+\beta V)\tan(V) -\beta\ln
\left( \frac{\frac{\pi}{2}W\cos(V)}{\frac{\pi}{2}+\beta V}
\right) \right] \nonumber \\
& +  & B_{1,\beta,\sigma,\mu},
\label{recipe2}
\end{eqnarray}
with
\begin{equation}
B_{1,\beta,\sigma,\mu}=\mu+\frac{2}{\pi}\beta\sigma\ln(\sigma).
\label{bb1}
\end{equation}
In the above equations $V$ and $W$ are independent random variables, such that
$V$ is uniformly distributed in the interval
$(-\frac{\pi}{2},\frac{\pi}{2})$ and
$W$ is exponentially distributed with a unit mean \cite{weron}. Numerical
integration has been performed for $\mu=0$ with increments of $\Delta
L_{\alpha,\beta}$ sampled from the strictly stable distributions \cite{janicki}, i.e. in general stable distributions generated \textit{via} Eqs.~(\ref{recipe1})--(\ref{bb1}) the value of $\mu$ is shifted
such that effectively $B_{\alpha,\beta,\sigma,\mu}$ is equal to 0.
Due to documented \cite{janicki}
numerical instability of all calculations with $\alpha$-stable random variables
 for the set $(\alpha=1,\beta\neq 0)$, this case has been excluded from the analysis.
Algorithms as described above have been positively tested for independence
on the choice of the integration step
(the evaluation of stochastic integrals with time steps $dt=10^{-4}$ and
$dt=10^{-5}$  yielded the same results).


\end{document}